\documentclass[letterpaper,aps,prd,twocolumn,tightenlines,preprintnumbers,nofootinbib,showkeys,superscriptaddress,longbibliography]{revtex4-1}
\pdfoutput=1

\usepackage{XCharter}
\usepackage{mathptmx}
\usepackage[T1]{fontenc}
\usepackage{blkarray}
\usepackage{fullpage}
\usepackage{amsfonts}
\usepackage{amsmath}
\usepackage{slashed}
\usepackage{amssymb}
\usepackage{graphicx}
\usepackage{epic}
\usepackage{eepic}
\usepackage{epsfig}
\usepackage{latexsym}
\usepackage[dvipsnames]{xcolor}
\usepackage[export]{adjustbox}
\usepackage{float}
\usepackage{multirow}
\usepackage{booktabs}
\usepackage{hyperref}
\usepackage{enumitem}
\hypersetup{colorlinks=true,citecolor=red,linkcolor=NavyBlue,urlcolor=NavyBlue}

\usepackage[caption=false]{subfig} 

\usepackage{natbib}
\usepackage{relsize}
\usepackage[left=1.6cm,right=1.6cm,top=1.65cm,bottom=1.65cm]{geometry}
\usepackage{comment}
\usepackage{shorthand}
\usepackage{dcolumn}
\usepackage{pifont}

\begin{document}

\title{Non-standard decays of vector-like top partners in a $2$-Higgs doublet model at the HL-LHC}

\author{Tanumoy Mandal}
\email{tanumoy@iisertvm.ac.in}
\affiliation{Indian Institute of Science Education and Research
             Thiruvananthapuram, Vithura, Kerala 695\,551, India}

\author{Stefano Moretti}
\email{stefano.moretti@cern.ch}
\affiliation{School of Physics and Astronomy, University of Southampton,
             Southampton SO17\,1BJ, United Kingdom}
\affiliation{Department of Physics and Astronomy, Uppsala University,
             Box 516, SE-751\,20 Uppsala, Sweden}

\author{Rachit Sharma}
\email{rachit21@iisertvm.ac.in}
\affiliation{Indian Institute of Science Education and Research
             Thiruvananthapuram, Vithura, Kerala 695\,551, India}

\begin{abstract}\noindent
Extensions of the Standard Model  featuring both an enlarged scalar sector
and vector-like fermions arise naturally in a wide class of well-motivated
theoretical frameworks. In such scenarios, vector-like Quarks (VLQs) can exhibit
non-standard decay modes involving additional Higgs states, giving rise to
distinctive collider signatures that remain largely unexplored by existing
experimental searches. We investigate the prospects of probing this
possibility at the high-luminosity Large Hadron Collider (HL-LHC) through the decay of vector-like top partner ($T$) to charged Higgs ($H^{\pm}$) followed by the decay, $H^\pm\to\tau\nu$, producing a final state
containing two tau leptons, two $b$-jets, and missing transverse energy.
A model-independent collider analysis is performed using global kinematic
observables constructed from visible objects and the missing transverse
momentum vector to suppress the dominant backgrounds.
Polarization-sensitive observables built from the hadronic $\tau$ decay
products are also examined as complementary probes of the 
spin-$0$ origin
of the $\tau$ leptons. The expected discovery sensitivity is evaluated using the
Asimov significance for an integrated luminosity of
$3~\mathrm{ab}^{-1}$ at $\sqrt{s}=14$~TeV. Our results demonstrate that
the $2\tau+2b+\slashed{E}_T$ (where $\slashed{E}_T$ is the missing transverse energy) channel provides a promising and largely
orthogonal avenue to search for non-standard VLQ decays in
extended Higgs sectors, with discovery-level sensitivity achievable for
VLQ masses up to approximately $1.9$~TeV.
\end{abstract}

\maketitle

\section{Introduction}
\label{sec:intro}

The discovery of the Higgs boson at the Large Hadron Collider
(LHC)~\cite{ATLAS:2012yve,CMS:2012qbp} has established the
Brout--Englert--Higgs mechanism as the origin of electroweak symmetry breaking (EWSB). Yet the precise structure of the Higgs sector remains only
partially explored, and there is no compelling theoretical reason for
it to be minimal. Extensions of the Standard Model (SM)
Higgs sector provide a well-motivated framework for probing deviations
from the SM predictions and for addressing open questions, such as the origin
of fermion mass hierarchies, the stability of the electroweak vacuum,
and the possibility of a strong first-order electroweak phase
transition relevant to baryogenesis.

Among the simplest and most thoroughly studied SM extensions is the $2$-Higgs Doublet Model (2HDM)~\cite{Gunion:1989we,Branco:2011iw}, which augments the SM by a second $SU(2)_L$ doublet while preserving consistency with the precision data through the alignment limit~\cite{Carena:2013ooa}. Beyond the SM-like Higgs boson $h$, the physical spectrum of the 2HDM contains two additional neutral states—a
CP-even scalar $H$ and a CP-odd pseudoscalar $A$—and a pair of charged
Higgs bosons $H^\pm$. The charged Higgs state is of particular
phenomenological interest because it has no SM analogue and its
observation would constitute unambiguous evidence for physics beyond the SM (BSM).

Vector-like quarks (VLQs) are a theoretically natural and well-motivated extension of the SM fermion sector. Their gauge-invariant masses are not generated by Yukawa couplings and are therefore free from the fine-tuning constraints that afflict chiral fermion masses. VLQs arise ubiquitously in many extensions of the SM such as composite Higgs models, little Higgs models, extra-dimensional scenarios etc.~\cite{Okada:2012gy,Gopalakrishna:2011ef,Gopalakrishna:2013hua,Aguilar-Saavedra:2013qpa,Ellis:2014dza,Alves:2023ufm,Banerjee:2024zvg},
and their mixing with the SM quarks can substantially modify the phenomenology of the Higgs sector. Importantly, when VLQs couple to an extended Higgs sector, new decay channels open up that have no analogue in the minimal setup~\cite{Gopalakrishna:2015wwa,Serra:2015xfa,
Dobrescu:2016pda,Arhrib:2017wmo,Aguilar-Saavedra:2017giu,
Chala:2017xgc,Moretti:2017qby,Bizot:2018tds,Colucci:2018vxz,
Han:2018hcu,Benbrik:2019zdp,Cacciapaglia:2019zmj,Xie:2019gya,
Dermisek:2019vkc,Wang:2020ips,Dermisek:2020gbr,Das:2020ozo,
Dermisek:2021zjd,Corcella:2021mdl,Dasgupta:2021fzw,Banerjee:2022izw,Bhardwaj:2022nko,Bhardwaj:2022wfz,Bardhan:2022sif,
Arhrib:2024tzm,Verma:2024kdx,Bardhan:2025pmn}. Such exotic VLQ decays---proceeding through additional neutral or charged
Higgs states---have been the subject of growing theoretical
interest~\cite{Angelescu:2015uiz,Benbrik:2015fyz,Dolan:2016eki,Arhrib:2016rlj,Cingiloglu:2023ylm}, yet current experimental searches remain largely insensitive to them.

Current LHC searches constrain the mass of the lightest VLQ partner of the top-quark (denoted by $T$) up to
approximately $1.3$--$1.5$~TeV under SM-like decay
assumptions~\cite{ATLAS:2022ozf,CMS:2022yxp}, i.e.,  decays into $Wb$, $Zt$, and $ht$. When such a VLQ couples
appreciably to a charged scalar, the exotic decay mode $T\to H^\pm b$
can become comparable to or dominant over the SM-like channels across
broad regions of parameter space, rendering the above exclusion limits
inapplicable. This motivates a dedicated study of VLQ signatures (of a top-quark companion)  
mediated by extended Higgs sector states. (See \cite{Benbrik:2019zdp} for the case of neutral Higgs bosons.)

\begin{figure}[b]
  \centering
  \includegraphics[width=\columnwidth]{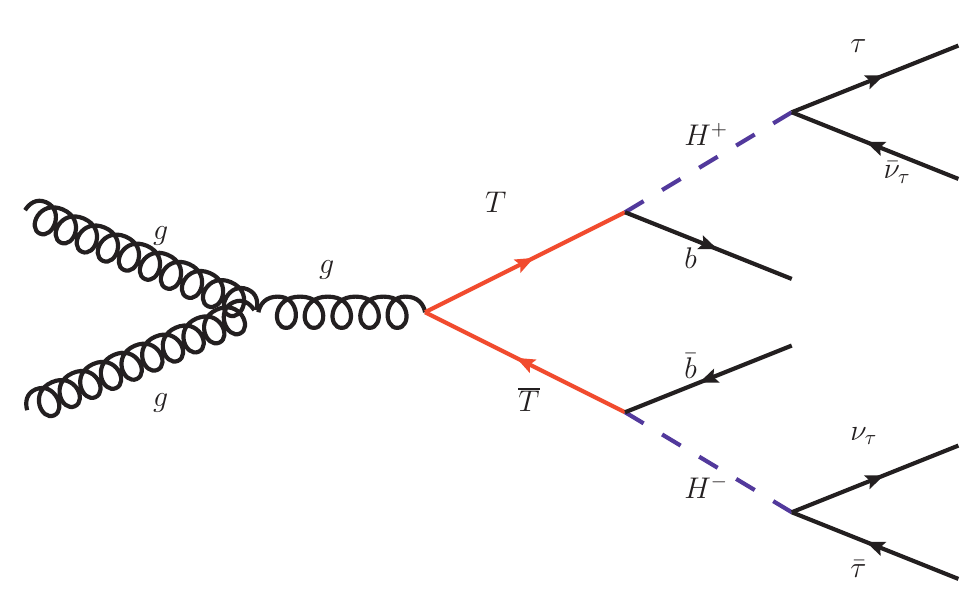}
  \caption{Representative Feynman diagram for the signal process
  $pp\to T\bar{T}\to(H^+b)(H^-\bar{b})\to(\tau^+\nu_\tau b)
  (\tau^-\bar{\nu}_\tau\bar{b})$, yielding the
  $2\tau+2b+\slashed{E}_T$ final state studied in this work.}
  \label{fig:Feyn}
\end{figure}

In this work, we investigate the $2\tau+2b+\slashed{E}_T$ final state
arising from pair-produced $T$ in a type-II 2HDM. The topology proceeds
through the cascade decay,
$T\bar{T}\to(H^+b)(H^-\bar{b})\to(\tau^+\nu_\tau b)(\tau^-\bar{\nu}_\tau\bar{b})$
(hereafter $\nu_\tau\equiv\nu$), exploiting the
enhanced $H^\pm\to\tau\nu$ branching ratio (BR) characteristic of large
$\tan\beta$ (defined in Sec.~\ref{sec:model}) in type-II models. The
analysis is formulated in a model-independent manner. The signal yield
is factorized as the product of the $pp\to T\bar{T}$ production cross
section and the effective cascade decay rate
$\mathcal{B}\equiv\mathrm{BR}(T\to H^\pm b)\times\mathrm{BR}(H^\pm\to\tau\nu)$,
so that results can be interpreted in any scenario involving a heavy
colored fermion decaying through a charged Higgs state. The signal
process is illustrated in Fig.~\ref{fig:Feyn}.

We assess the discovery and exclusion reach of the high-luminosity LHC (HL-LHC)
($\mathcal{L}=3~\mathrm{ab}^{-1}$, $\sqrt{s}=14$~TeV) through a
cut-based analysis employing global kinematic observables. The complementary role of $\tau$ polarization properties in distinguishing the signal from the SM backgrounds is also discussed.

The paper is organized as follows. The theoretical framework is
described in Sec.~\ref{sec:model}. The collider analysis strategy is
presented in Sec.~\ref{sec:analysis}. The $\tau$ lepton polarization observables are
discussed in Sec.~\ref{sec:pol}. The main results are presented in
Sec.~\ref{sec:results} while conclusions are given in
Sec.~\ref{sec:conclusion}.

\section{Theoretical Framework}
\label{sec:model}

\subsection{The 2HDM}
\noindent
We consider a CP-conserving 2HDM with a softly broken discrete
$\mathbb{Z}_2$ symmetry. The scalar potential involving the two
$SU(2)_L$ doublets $(\Phi_1,\Phi_2)$ reads~\cite{Branco:2011iw,Gunion:1989we},
\begin{eqnarray}
V(\Phi_1,\Phi_2) &=&
m_{11}^2\,\Phi_1^\dagger\Phi_1 + m_{22}^2\,\Phi_2^\dagger\Phi_2
- m_{12}^2\!\left(\Phi_1^\dagger\Phi_2+\mathrm{h.c.}\right) \nonumber\\
&+&\frac{\lambda_1}{2}(\Phi_1^\dagger\Phi_1)^2
+\frac{\lambda_2}{2}(\Phi_2^\dagger\Phi_2)^2
+\lambda_3(\Phi_1^\dagger\Phi_1)(\Phi_2^\dagger\Phi_2)\nonumber\\
&+&\lambda_4(\Phi_1^\dagger\Phi_2)(\Phi_2^\dagger\Phi_1)
+\frac{\lambda_5}{2}\!\left[(\Phi_1^\dagger\Phi_2)^2+\mathrm{h.c.}\right],
\label{eq:thdmV}
\end{eqnarray}
where all parameters are real. Each doublet acquires
a VEV, $\langle\Phi_i\rangle=(0,\,v_i/\sqrt{2})^T$ for $i=1,2$, with
$v_1$ and $v_2$ the VEVs of $\Phi_1$ and $\Phi_2$, respectively.
Working in the Higgs basis $(H_1,H_2)$
in which only one doublet acquires a non-zero vacuum expectation value
(VEV),
\begin{equation}
H_1=\begin{pmatrix}G^+\\\dfrac{v+\varphi_1^0+iG^0}{\sqrt{2}}\end{pmatrix},
\quad
H_2=\begin{pmatrix}H^+\\\dfrac{\varphi_2^0+iA}{\sqrt{2}}\end{pmatrix},
\end{equation}
where $G^{0,\pm}$ are the Goldstone bosons absorbed by the $Z$ and
$W^\pm$ whereas $H^\pm$ is the physical charged Higgs state and $A$ is the CP-odd
neutral one. The physical CP-even states arise from the mixing
\begin{equation}
\begin{pmatrix}h\\H\end{pmatrix}
=\begin{pmatrix}
\sin(\beta-\alpha) & \cos(\beta-\alpha)\\
\cos(\beta-\alpha) & -\sin(\beta-\alpha)
\end{pmatrix}
\begin{pmatrix}\varphi_1^0\\\varphi_2^0\end{pmatrix},
\end{equation}
where $\tan\beta=v_2/v_1$ with $v_1$ and $v_2$ the VEVs of $\Phi_1$
and $\Phi_2$ defined above, with $v=\sqrt{v_1^2+v_2^2}=246$~GeV, and the mixing angle
$\alpha$ diagonalizes the CP-even mass matrix. In the alignment limit,
$\sin(\beta-\alpha)\to 1$, the lighter state $h$ reproduces the SM
Higgs couplings, consistent with current LHC
measurements~\cite{CMS:2022dwd,ATLAS:2022vkf}.
In the type-II 2HDM, down-type quarks and charged leptons couple
exclusively to $\Phi_1$ whereas up-type quarks couple to $\Phi_2$. As
a consequence, the $H^\pm\tau\nu$ Yukawa coupling is proportional to
$m_\tau\tan\beta$, making $\mathrm{BR}(H^\pm\to\tau\nu)$ particularly
large at high $\tan\beta$—a regime that is phenomenologically well
motivated and experimentally less constrained for $M_{H^\pm}$ above
the top-quark threshold.

\subsection{VLQs as top partners}

\noindent
We extend the type-II 2HDM by a singlet up-type VLQ pair
$(T_L,T_R)$ with electric charge $Q=2/3$ (henceforth, top--VLQ). The most general
renormalizable Yukawa Lagrangian involving the third-generation quark
sector is
\begin{align}
-\mathcal{L}_Y^{\mathrm{II}}\supset
\Big\{&\;y_t\,\overline{Q_L^0}\,\widetilde{H}_2\,t_R^0
+y_T\,\overline{Q_L^0}\,\widetilde{H}_2\,T_R^0
+\xi_T\,\overline{Q_L^0}\,\widetilde{H}_1\,T_R^0\nonumber\\
&+\widetilde{\omega}_T M_T\,\overline{T_L^0}\,t_R^0
+M_T\,\overline{T_L^0}T_R^0+\mathrm{h.c.}\Big\},
\label{eq:LagVLQgen}
\end{align}
where $Q_L^0=(t_L^0,b_L^0)^T$, $t_R^0$ is the
third-generation SM up-type singlet, and
$\widetilde{H}_i=i\tau_2H_i^*$. The off-diagonal mass term
$\widetilde{\omega}_T$ can be removed by a field redefinition in the
right-handed sector, leaving Eq.~\eqref{eq:LagVLQgen} with adjusted 
parameters~\cite{Dubey:2025sfh}. After EWSB, the top--VLQ mass
matrix in the weak-interaction basis $(t_L^0,T_L^0)$ is
\begin{equation}
\mathcal{M}=
\begin{pmatrix}
\dfrac{y_t v}{\sqrt{2}} & \dfrac{\xi_T v}{\sqrt{2}}\\[6pt]
0 & M_T
\end{pmatrix},
\label{eq:massmat}
\end{equation}
which is diagonalized by a bi-unitary transformation
$U_L\,\mathcal{M}\,U_R^\dagger=\mathcal{M}_d$, yielding the physical
mass eigenstates
\begin{equation}
\begin{pmatrix}t\\T\end{pmatrix}_{L,R}
=\begin{pmatrix}
\cos\theta_{L,R} & -\sin\theta_{L,R}\\
\sin\theta_{L,R} & \phantom{-}\cos\theta_{L,R}
\end{pmatrix}
\begin{pmatrix}t^0\\T^0\end{pmatrix}_{L,R}.
\end{equation}
The left- and right-handed mixing angles are given by,
\begin{align}
\tan(2\theta_L)&=\frac{4m_t M_T}{2M_T^2-2m_t^2-\xi_T^2v^2},
\label{eq:thetaL}\\
\tan(2\theta_R)&=\frac{2\sqrt{2}\,\xi_T m_t v}
{2M_T^2+2m_t^2-\xi_T^2v^2}.
\label{eq:thetaR}
\end{align}
After diagonalization, the physical $T$ state couples to the charged
Higgs boson through a mixing-angle-suppressed $TbH^\pm$ vertex. The
standard decay channels $T\to Wb$, $T\to Zt$, $T\to ht$ are always
present. The same mixing also generates the non-standard neutral-scalar
modes $T\to Ht$ and $T\to At$, which compete with $T\to H^\pm b$ for
branching ratio. However, these neutral-scalar channels do not
contribute to the $2\tau+2b+\slashed{E}_T$ final state under study: the
subsequent top decay $t\to Wb$ introduces additional jets or leptons
from the $W$, populating busier final states distinct from our signal
region. Any such competition for the BR is automatically
absorbed into our model-independent treatment, since
$\mathrm{BR}(T\to H^\pm b)$ is kept as a free parameter within
$\mathcal{B}$ rather than fixed to unity except in the optimistic
$\mathcal{B}=1$ benchmark of Sec.~\ref{sec:results}. The non-standard
mode $T\to H^\pm b$ can become the dominant decay for sizable $\xi_T$
and moderate $M_{H^\pm}/M_T$ ratio.

\subsection{Model-independent signal parametrization}

\noindent
The signal cross section in the $2\tau+2b+\slashed{E}_T$ channel is
\begin{equation}
\sigma_{\mathrm{sig}}=\sigma(pp\to T\bar{T})
\times\mathrm{BR}(T\to H^\pm b)^2
\times\mathrm{BR}(H^\pm\to\tau\nu)^2.
\label{eq:sigyield}
\end{equation}
Since the intermediate particles are not individually reconstructible
at the detector level, the analysis is sensitive only to the product
$\mathcal{B}^2\equiv[\mathrm{BR}(T\to H^\pm b)\times
\mathrm{BR}(H^\pm\to\tau\nu)]^2$. We therefore present results as
contours of constant Asimov significance in the planes of
$\{M_T,\mathcal{B}\}$ and $\{M_{H^\pm},M_T\}$ (at $\mathcal{B}=1$),
enabling reinterpretation in any scenario in which a heavy
color-triplet fermion ($Q$) undergoes the specific cascade
$Q\to H^\pm q \to \tau\nu\, q$. The leptonic decay
$H^\pm\to\tau\nu$ is a distinctive feature of the type-II 2HDM at
large $\tan\beta$, where the $H^\pm\tau\nu$ Yukawa coupling is
enhanced by $m_\tau\tan\beta$, and this mode dominates over
$H^\pm\to tb$ once $\tan\beta\gtrsim 2$--$3$ for charged Higgs masses
above the top-quark threshold~\cite{Branco:2011iw}. The same topology
can arise in other extended scalar sectors with a similar Yukawa
structure, such as the type-$X$ or flipped 2HDM variants, or in
supersymmetric models where the lightest chargino decays through a
stau. In scenarios where $H^\pm\to tb, W^{(*)}h$ and/or $W^{(*)}A$ dominate instead, the present results do not directly apply, and a dedicated analysis in those decay channels would be required.

\section{Collider Analysis}
\label{sec:analysis}

\subsection{Signal and background processes}

\noindent
The signal topology we consider here is as follows,
\begin{equation}
pp\to T\bar{T}\to(H^+b)(H^-\bar{b})\to
(\tau^+\nu_\tau\,b)(\tau^-\bar{\nu}_\tau\,\bar{b}).
\end{equation}
This yields the $2\tau+2b+\slashed{E}_T$ final state (Fig.~\ref{fig:Feyn}).
Both tau leptons are required to decay hadronically
($\tau\to\mathrm{hadrons}+\nu_\tau$), giving two reconstructed
hadronic tau candidates $\tau_h$ in the detector. This mode has a
combined BR of approximately $42\%$ and benefits from
efficient hadronic tau identification and $b$-tagging at the LHC detectors.
Signal samples are generated for a grid of benchmark mass points
spanning $M_T\in[600,2200]$~GeV and $M_{H^\pm}\in[200,800]$~GeV.
The $T\bar{T}$ production cross sections are evaluated at next-to-leading order (NLO) in QCD,
consistent with available higher order 
calculations~\cite{Aliev:2010zk}.

The dominant SM background processes considered are:
\begin{itemize}
\item $t\bar{t}+$jets, where each top decays as $t\to Wb$ with both
$W$ bosons decaying to $\tau\nu$, directly supplying the two $b$-jets,
the two taus, and the missing energy that define the signal topology;
this is the dominant background in the analysis.
\item $t\bar{t}V$ associated production ($V=W,Z,H$), where the
$t\bar{t}$ system already supplies the $2\tau+2b+\slashed{E}_T$
topology while $V$ contributes additional jets or missing energy;
this constitutes the second-largest background.
\item Single vector boson production ($Z\to\tau^+\tau^-$ or
$W\to\tau\nu$) accompanied by jets. Despite the large inclusive cross
section, the requirement of two genuine $b$-tagged jets strongly
suppresses this background, owing to the small $b$-tagging and
tau-misidentification rates for light- and heavy-flavor jets; this
background is found to be negligible after selection.
\item $tW$ associated single-top production, where the $b$-jet from
top decay and a second $b$-jet from gluon splitting, together with
$\tau\nu$ from the $W$ decays, can mimic the signal topology at a
smaller rate than $t\bar{t}$.
\item $tb$ associated single-top production, which naturally supplies
two $b$-jets but only a single tau from $t\to Wb\to(\tau\nu)b$, making
it subdominant relative to $t\bar{t}$ and $tW$.
\item Diboson production ($WW$, $ZZ$) with jets, where
$WW\to\tau^+\nu_\tau\,\tau^-\bar\nu_\tau$ or
$ZZ\to\tau^+\tau^-\,\nu\bar\nu$ directly yield the $2\tau+\slashed{E}_T$
topology, with the two $b$-jets arising from gluon splitting. $WZ+$jets
is omitted since it cannot efficiently populate the exactly-two-$\tau_h$
signal region: the accompanying $W$ decay yields either an extra
isolated lepton (rejected by the lepton veto) or a third tau (rejected
by the exactly-two-$\tau_h$ requirement).
\end{itemize}

The higher-order cross sections used to derive $K$-factors for each
background are listed in Table~\ref{tab:Backgrounds}. All backgrounds
are normalized to these higher-order predictions throughout the
analysis. The $K$-factors account for the ratio of the higher-order
cross sections to the corresponding LO predictions and are
applied uniformly after event selection.

\begin{table}[!htbp]
\caption{Higher-order inclusive cross sections for the dominant SM
background processes at $\sqrt{s}=14$~TeV, prior to decay and event
selection. The QCD accuracy of each calculation is indicated in the
final column. These values are used to derive the $K$-factors applied
to the cross section obtained at the leading-order (LO).}
\label{tab:Backgrounds}
\centering
\renewcommand{\arraystretch}{1.35}
\begin{tabular*}{\columnwidth}{l @{\extracolsep{\fill}} l r c}
\toprule
\multicolumn{2}{l}{Background process} & $\sigma\,(\mathrm{pb})$ & Order\\
\midrule
\multirow{2}{*}{$VV+$jets~\cite{Campbell:2011bn}}
  & $WW+$jets & $124.31$ & NLO\\
  & $ZZ+$jets & $17.72$  & NLO\\
\midrule
Single top~\cite{Kidonakis:2015nna}
  & $tW$      & $83.10$  & N$^2$LO\\
\midrule
$t\bar{t}$~\cite{Muselli:2015kba}
  & $t\bar{t}+$jets & $988.57$ & N$^3$LO\\
\midrule
\multirow{3}{*}{$t\bar{t}V$~\cite{Kulesza:2018tqz,Balsach:2025jcw}}
  & $t\bar{t}Z$ & $1.05$ & NLO+N$^2$LL\\
  & $t\bar{t}W$ & $0.65$ & NLO+N$^2$LL\\
  & $t\bar{t}H$ & $0.64$ & NNLO+NNLL+EW\\
\bottomrule
\end{tabular*}
\end{table}

\subsection{Event generation and preselection cuts}

Signal and background events are generated with
\textsc{MadGraph5\_aMC@NLO}~\cite{Alwall:2014hca} at LO accuracy in QCD, interfaced with
\textsc{Pythia~8}~\cite{Sjostrand:2014zea} for parton showering,
hadronization, and underlying-event modelling.
Jets are clustered with the anti-$k_T$
algorithm~\cite{Cacciari:2008gp} using radius parameter $R=0.4$ as
implemented in \textsc{FastJet}~\cite{Cacciari:2011ma}. Detector
effects are modelled with \textsc{Delphes3}~\cite{deFavereau:2013fsa}
using the ATLAS detector card.

The baseline object selection requires exactly two hadronic tau candidates ($\tau_h$) satisfying $p_T>30$~GeV and
$|\eta|<2.5$, and at least two $b$-tagged jets with $p_T>30$~GeV and
$|\eta|<2.5$. Tau candidates overlapping with $b$-jets within
$\Delta R<0.4$ are discarded. Events containing isolated electrons or
muons with $p_T>20$~GeV are vetoed to suppress the dileptonic
$t\bar{t}$ contribution.

\subsection{Kinematic observables and event selection}

Invariant-mass variables involving missing transverse energy require
special care at a hadron collider, since the longitudinal momentum of
the neutrino system is unmeasured and a fully Lorentz-invariant
four-momentum cannot be assigned to $\slashed{E}_T$ on an event-by-event
basis. We resolve this by constructing an effective four-vector,
\begin{equation}
p^\mu_{\slashed{E}_T}=\bigl(\slashed{E}_x,\;\slashed{E}_y,\;0,\;
\slashed{E}_T\bigr),
\label{eq:METvec}
\end{equation}
where $\slashed{E}_{x,y}$ are the Cartesian components of the missing
momentum and $\slashed{E}_T=|\vec{\slashed{p}}_T|$. This vector is
combined with the four-momenta of visible objects to define multi-body
effective invariant masses
$\mathrm{IM}(\slashed{E}_T,X_1,X_2,\ldots)$. Although these
quantities are not strictly Lorentz-invariant, they are well-defined
and experimentally computable, and are found to provide strong
signal-to-background discrimination.

The primary kinematic discriminants employed in the analysis are:
\begin{itemize}
\item $\slashed{E}_T$: missing transverse energy, sensitive to the
      energetic neutrinos from $H^\pm\to\tau\nu$;
\item $S_T$: scalar sum of the transverse momenta of all reconstructed
      objects including $\slashed{E}_T$, characterizing the overall
      event energy scale;
\item $\mathrm{IM}(\slashed{E}_T,\tau_{j1},\tau_{j2})$: effective
      invariant mass of the two leading tau-jets combined with
      $\slashed{E}_T$;
\item $\mathrm{IM}(\slashed{E}_T,\tau_{j1},b_{j1})$ and
      $\mathrm{IM}(\slashed{E}_T,\tau_{j1},b_{j2})$: three-body
      invariant masses sensitive to the cascading mass scale of the
      signal;
\item $\mathrm{IM}(\slashed{E}_T,\tau_{j1})$: two-body transverse
      mass-like observable.
\end{itemize}

We emphasize that every invariant-mass variable of the form $\mathrm{IM}(\slashed{E}_T,\ldots)$ listed above, both in the selection criteria and in the distributions shown below, is constructed using the effective missing-momentum four-vector defined in Eq.~\eqref{eq:METvec}.

\begin{figure*}[t]
\centering
\includegraphics[width=0.32\textwidth]{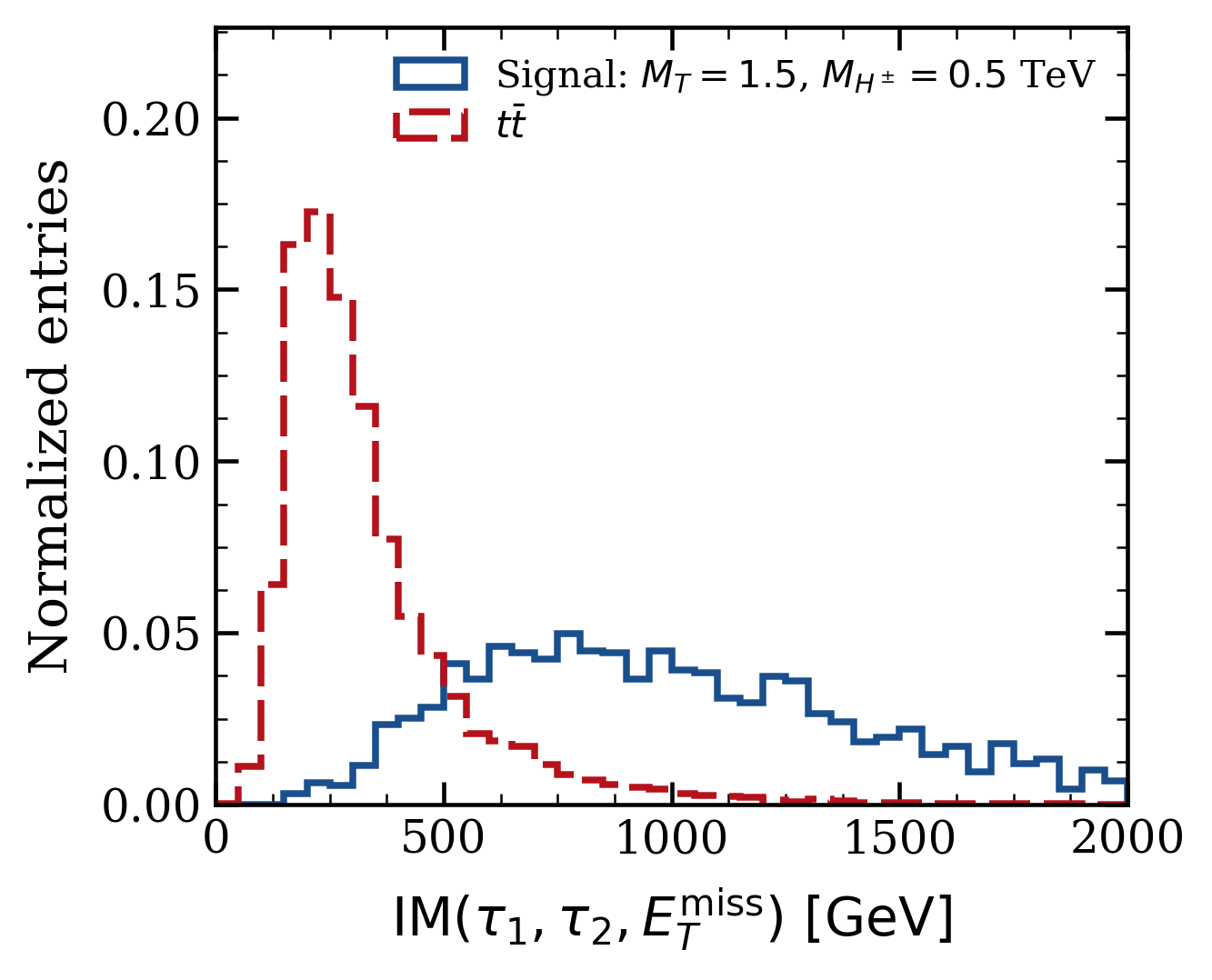}
\includegraphics[width=0.32\textwidth]{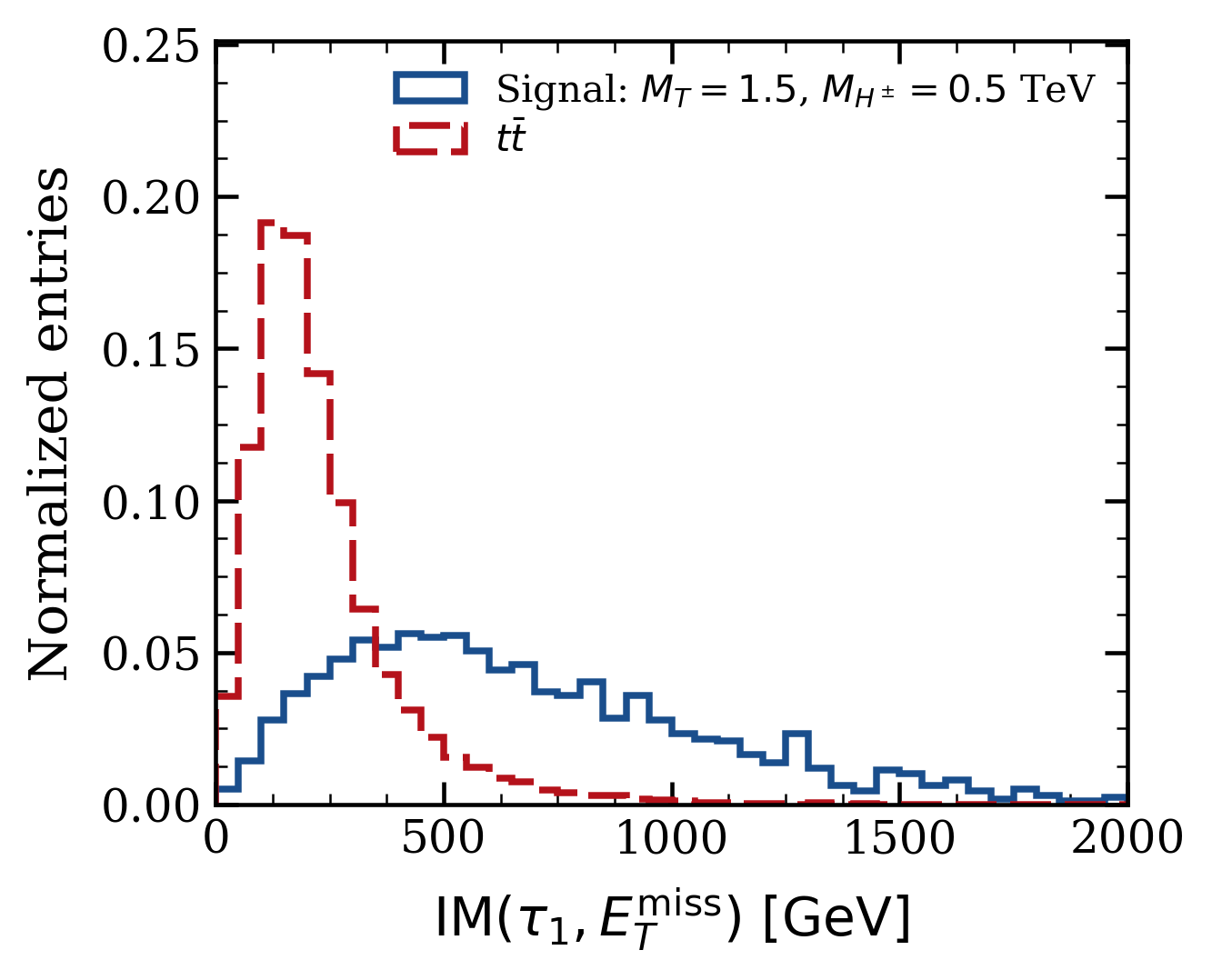}
\includegraphics[width=0.32\textwidth]{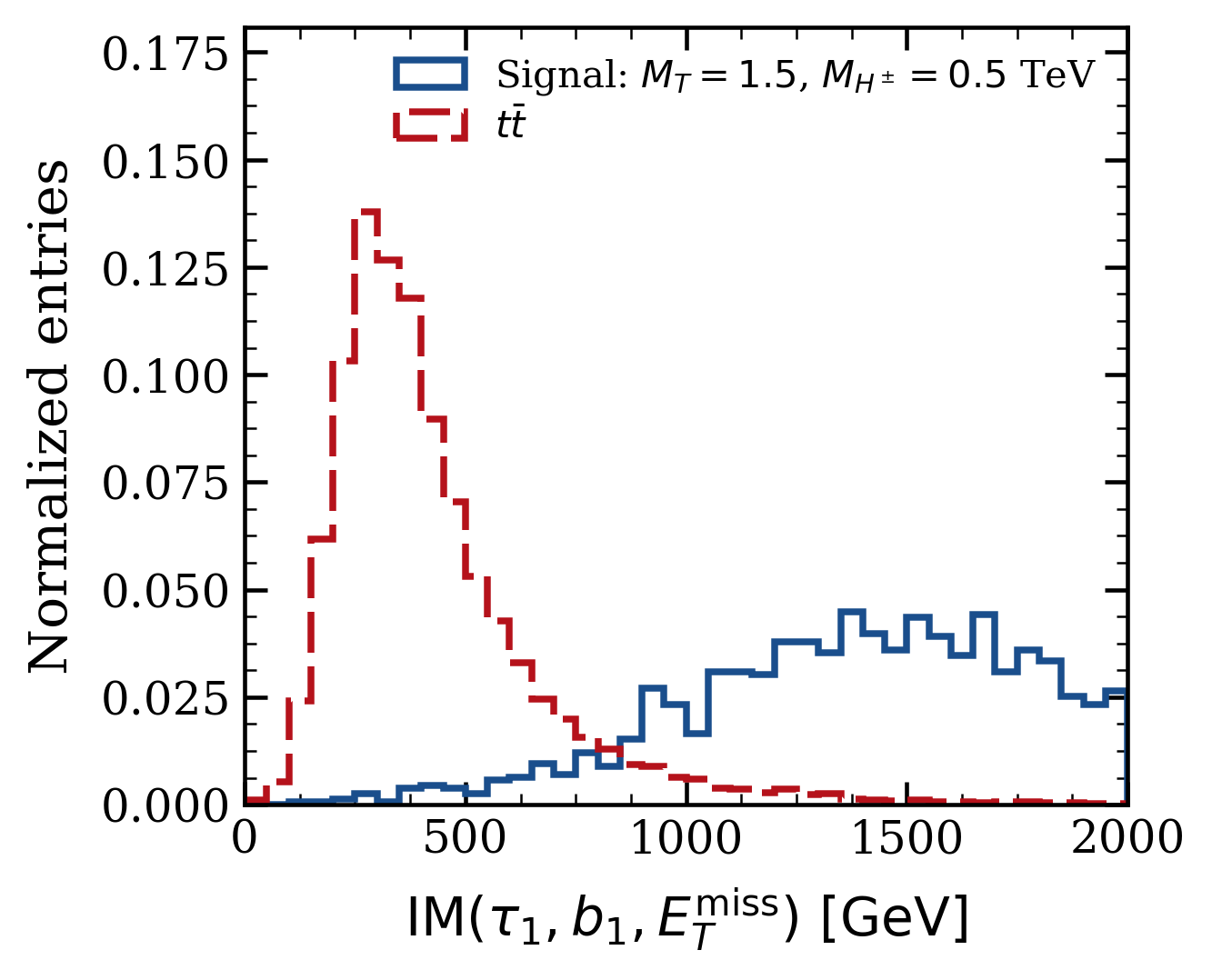}\\[4pt]
\includegraphics[width=0.32\textwidth]{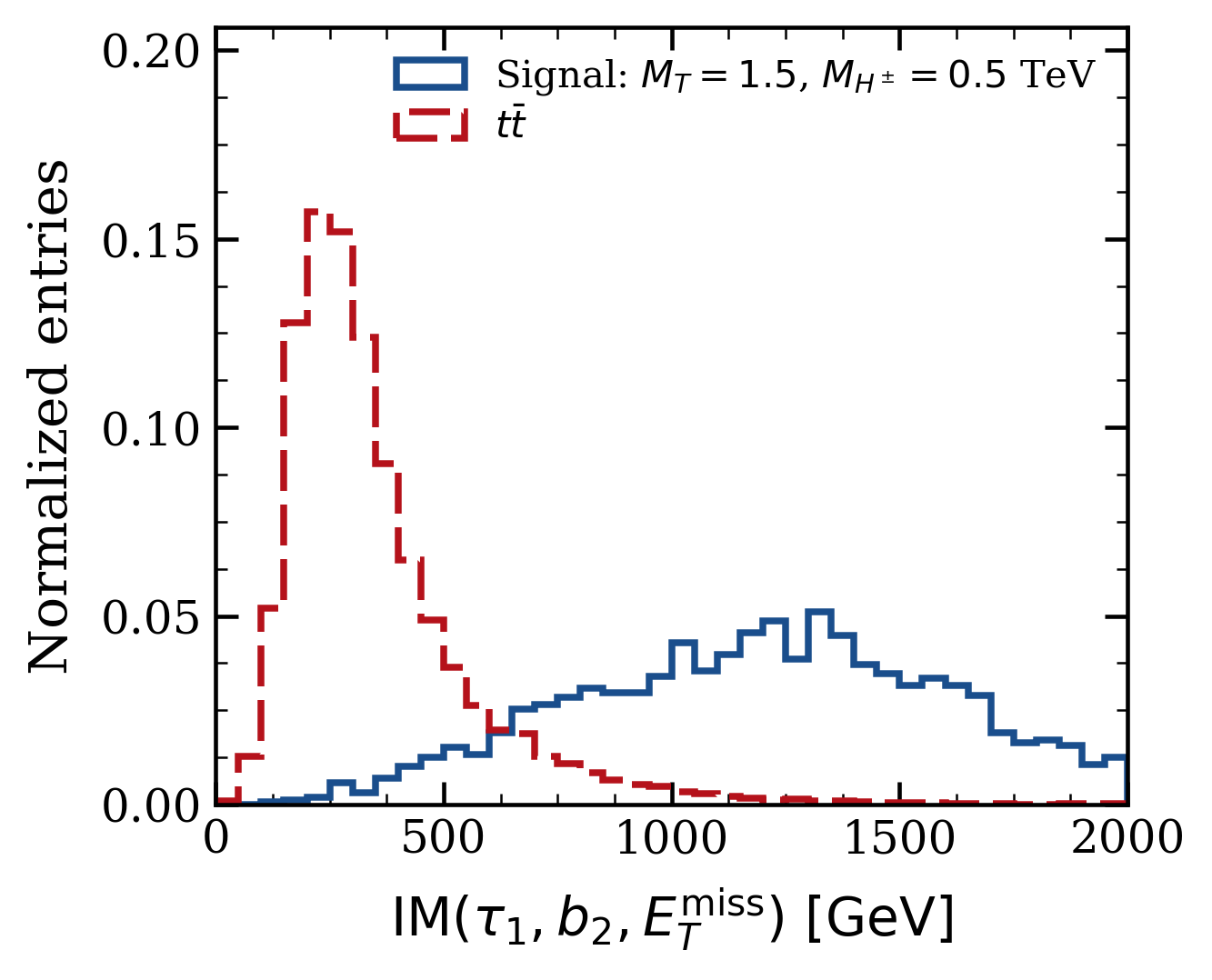}
\includegraphics[width=0.32\textwidth]{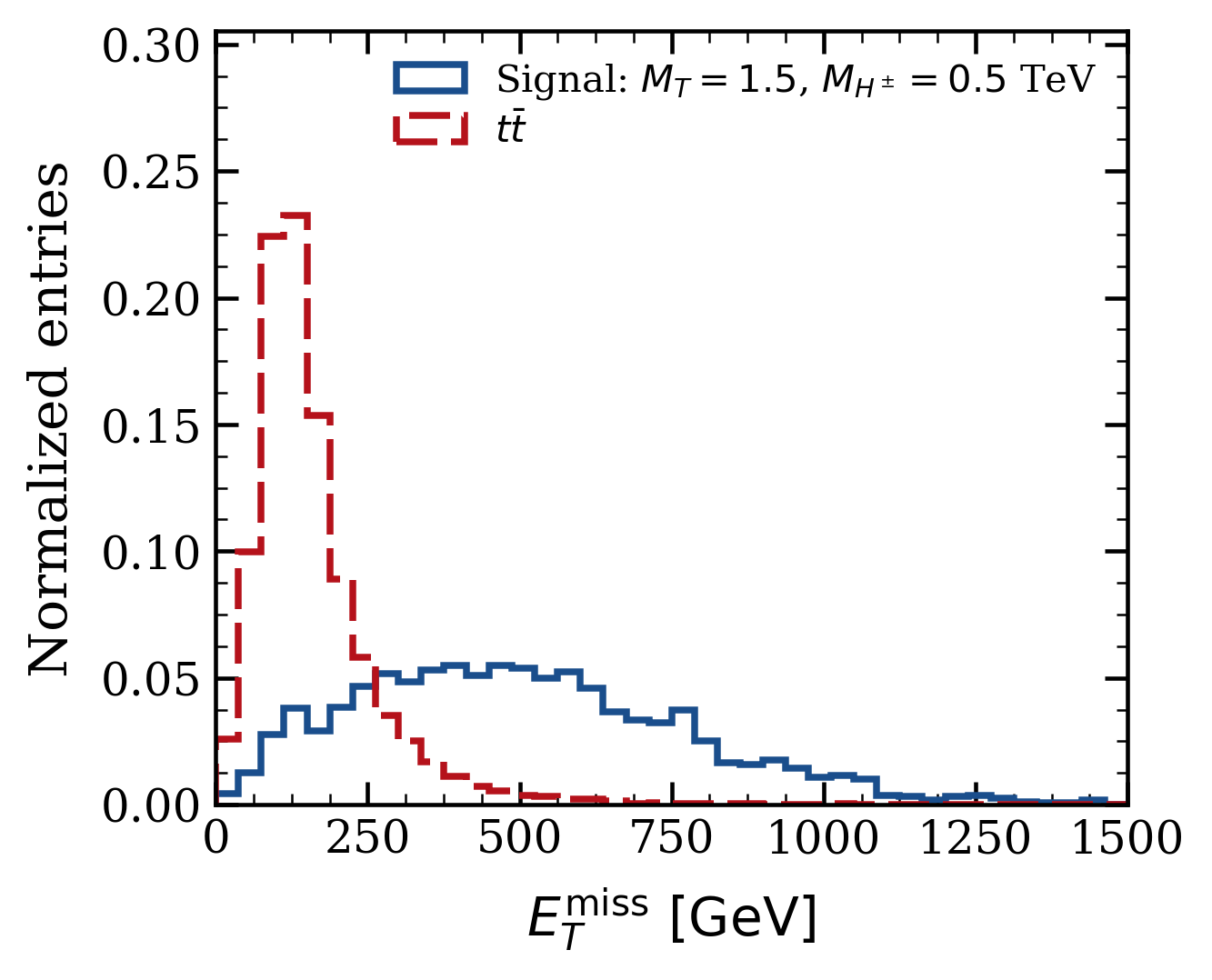}
\includegraphics[width=0.32\textwidth]{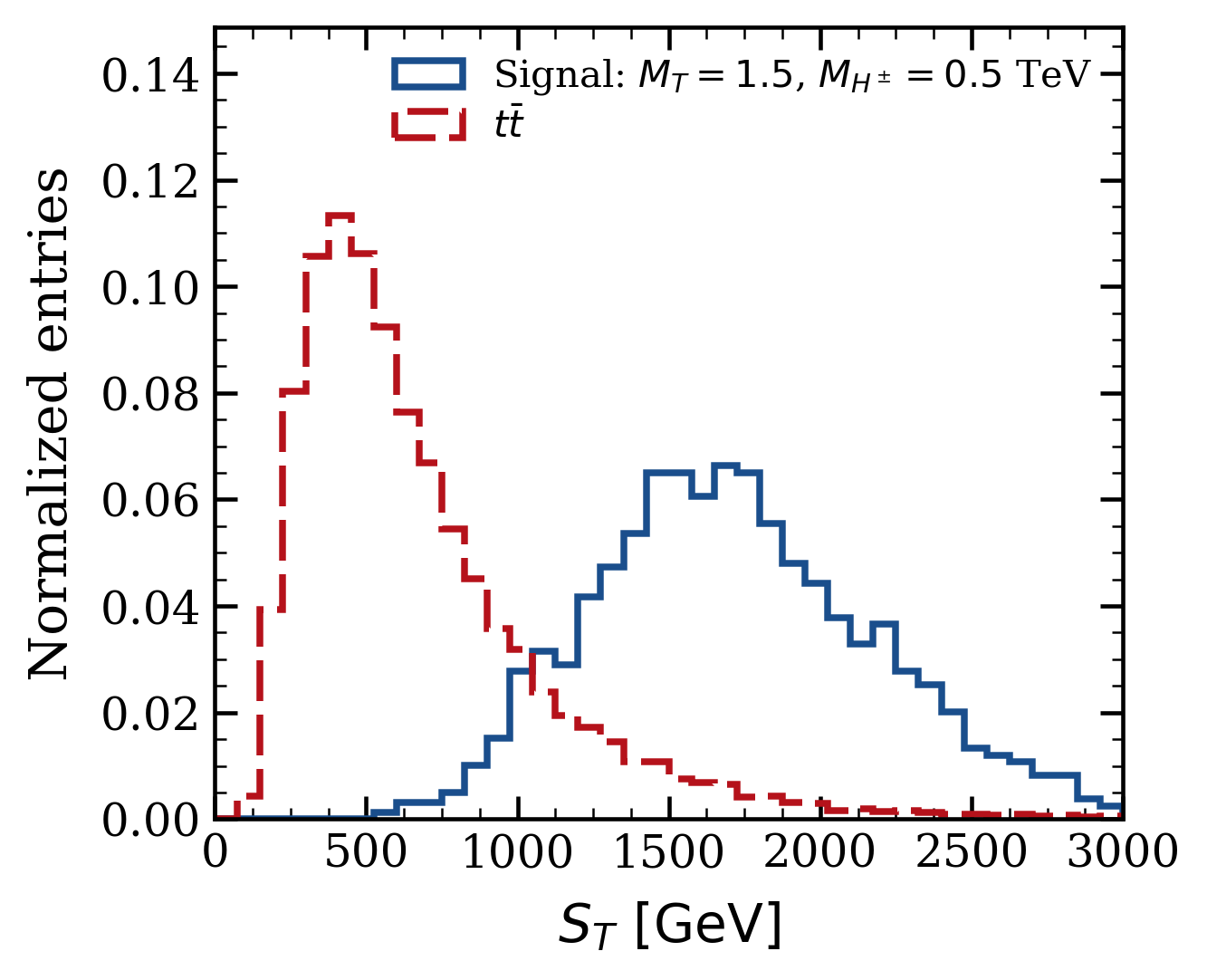}
\caption{Normalized distributions of the primary kinematic
discriminants for the signal benchmark with $M_T=1500$~GeV and
$M_{H^\pm}=500$~GeV, and for the dominant $t\bar{t}$ background.
\textit{Top row} (left to right):
$\mathrm{IM}(\slashed{E}_T,\tau_{j1},\tau_{j2})$,
$\mathrm{IM}(\slashed{E}_T,\tau_{j1})$, and
$\mathrm{IM}(\slashed{E}_T,\tau_{j1},b_{j1})$.
\textit{Bottom row}:
$\mathrm{IM}(\slashed{E}_T,\tau_{j1},b_{j2})$,
$\slashed{E}_T$, and $S_T$.
The missing transverse energy four-vector entering all
$\mathrm{IM}(\slashed{E}_T,\ldots)$ variables is constructed using the
effective four-vector of Eq.~\eqref{eq:METvec}.
Signal distributions exhibit considerably harder spectra than the
$t\bar{t}$ background across all observables, driven by the large
invariant mass scales of the cascade decay. \label{fig:kinematic_distributions}}
\end{figure*}

Normalized distributions of these observables are shown in
Fig.~\ref{fig:kinematic_distributions} for the signal benchmark with
$M_T=1500$~GeV, $M_{H^\pm}=500$~GeV, and for the dominant $t\bar{t}$
background. The signal exhibits consistently harder spectra across
all variables, driven by the large invariant mass scales $M_T$ and
$M_{H^\pm}$ of the cascade. This kinematic separation provides the
basis for the cut-based selection described below.

Based on the discriminating power of the observables described above,
a sequential cut-based selection is applied, optimized to maximize
the Asimov significance for a benchmark signal with $M_T=1500$~GeV
and $M_{H^\pm}=500$~GeV. The selection criteria are summarized in
Table~\ref{tab:cutflow}. The global $\slashed{E}_T$ and $S_T$
requirements (C1 and C2) efficiently reduce the multi-jet and low-mass
backgrounds, while the multi-body invariant-mass cuts (C3) further
exploit the distinctive heavy-particle topology of the signal. The
invariant-mass variables in C3 are constructed using the effective
four-vector of Eq.~\eqref{eq:METvec}.

\begin{table}[htbp]
\caption{Sequential event selection criteria applied in the cut-based
analysis. The invariant-mass variables involving $\slashed{E}_T$ are
constructed using the effective four-vector of
Eq.~\eqref{eq:METvec}. Cuts are ordered by their approximate
discriminating power, progressing from global energy-scale requirements
(C1, C2) to topology-specific invariant-mass requirements (C3).}
\label{tab:cutflow}
\renewcommand{\arraystretch}{1.35}
\begin{tabular*}{\columnwidth}{c @{\extracolsep{\fill}} l}
\toprule
Label & Selection criterion\\
\midrule
C1 & $\slashed{E}_T > 350~\mathrm{GeV}$\\
C2 & $S_T > 1200~\mathrm{GeV}$\\
\midrule
\multirow{4}{*}{C3}
  & $\mathrm{IM}(\slashed{E}_T,\tau_{j1},\tau_{j2})>500~\mathrm{GeV}$\\
  & $\mathrm{IM}(\slashed{E}_T,\tau_{j1},b_{j1})>1100~\mathrm{GeV}$\\
  & $\mathrm{IM}(\slashed{E}_T,\tau_{j1},b_{j2})>700~\mathrm{GeV}$\\
  & $\mathrm{IM}(\slashed{E}_T,\tau_{j1})>300~\mathrm{GeV}$\\
\bottomrule
\end{tabular*}
\end{table}

\begin{table}[htbp]
\caption{Cut-flow table showing the number of expected events
surviving each successive selection requirement for the dominant SM
backgrounds and for two representative signal benchmarks, normalized
to an integrated luminosity of $3~\mathrm{ab}^{-1}$ at
$\sqrt{s}=14$~TeV. All numbers are rounded to the nearest integer
except entries below unity, which are quoted to one decimal place.
The signal benchmark labels denote $(M_T,\,M_{H^\pm})$ in GeV.\label{tab:cutflow_events}}

\renewcommand{\arraystretch}{1.35}
\begin{tabular*}{\columnwidth}{l @{\extracolsep{\fill}} r r r}
\toprule
Process & C1 & C2 & C3 \\
\midrule
$t\bar{t}$                          & 1517  & 943  & 33   \\
$t\bar{t}H$                         & 3     & 1    & $<1$ \\
$t\bar{t}Z$                         & 11    & 3    & $<1$ \\
\midrule
Total background                    & 1531  & 947  & 33   \\
\midrule
Signal $(1500,\,500)$               & 160   & 142  & 94   \\
Signal $(2000,\,500)$               & 12    & 12   & 9    \\
\bottomrule
\end{tabular*}
\end{table}

\subsection{Statistical analysis}

The expected discovery significance is evaluated using the Asimov formalism~\cite{Cowan:2010js}. The Asimov $Z$-score is computed from the profile likelihood ratio including a Gaussian
constraint on the background normalization,
\begin{multline}
Z_{\mathrm{A}} =
\Bigg[
2(s+b)\ln\!\left(\frac{(s+b)(b+\sigma_b^2)}{b^2+(s+b)\sigma_b^2}\right)\\
-\frac{2b^2}{\sigma_b^2}\ln\!\left(1+\frac{s\sigma_b^2}{b(b+\sigma_b^2)}\right)
\Bigg]^{1/2}\!,
\label{eq:ZA}
\end{multline}
where $s$ and $b$ are the total expected numbers of signal and
background events surviving all selections, respectively, and $\sigma_b$ is the
absolute systematic uncertainty on the background normalization.
A flat uncertainty of $\sigma_b/b=10\%$ is assumed throughout,
reflecting the combined effect of theoretical uncertainties on the
background cross sections as well as experimental uncertainties on tau
identification and $b$-tagging efficiencies. A significance of
$Z_\mathrm{A}\geq5\,(2)$ is adopted as the discovery (exclusion)
threshold.

\section{Tau Polarization as a Complementary Probe}
\label{sec:pol}

The polarization state of the tau leptons encodes information about
the spin of their parent particle that is not captured by the
kinematic observables used in the event selection. In the present
signal, taus originate from the decay of the spin-0 charged Higgs
boson, $H^\pm\to\tau^\pm\nu_\tau$. The chiral structure of the
$H^\pm\tau\nu$ Yukawa coupling implies that $\tau^+$ ($\tau^-$)
produced in this decay are in right-handed (left-handed) helicity
eigenstates, i.e.,~$\mathcal{P}_\tau=+1$. This is opposite to the
case of $W^\pm\to\tau^\pm\nu_\tau$ decays via the transverse gauge
component, where the purely $V$-$A$ electroweak coupling forces
$\tau^+$ into a left-handed state ($\mathcal{P}_\tau=-1$). We note
that the longitudinal (Goldstone) component of the $W$ boson couples
to $\tau\nu$ through a scalar Yukawa-type vertex of the same chiral
structure as the $H^\pm\tau\nu$ coupling, and would therefore yield
$\mathcal{P}_\tau=+1$ like the signal. However, this contribution to
$\Gamma(W\to\tau\nu)$ is suppressed relative to the transverse gauge
contribution by $(m_\tau/M_W)^2\sim5\times10^{-4}$, and is therefore
entirely negligible. The dominant $t\bar{t}$ background thus produces
taus with $\mathcal{P}_\tau=-1$ via the transverse $W$ component,
opposite to the signal, providing a powerful handle for discrimination
that is absent from the kinematic selection of
Sec.~\ref{sec:analysis}. 

The polarization of the tau can be accessed experimentally through
the energy and angular distributions of its visible hadronic decay
products. For one-prong decays through the $\pi$ and $\rho$ modes,
we construct the polarization-sensitive variables $R_1$ and $R_2$,
defined as normalized ratios of visible momenta of the decay
products~\cite{Bullock:1992yt,Moretti:2002ny}. These variables are
directly accessible from the reconstructed constituents of the tau
jet, making them suitable for use in an experimental analysis.

\begin{figure*}[t]
\centering
\includegraphics[width=0.30\textwidth]{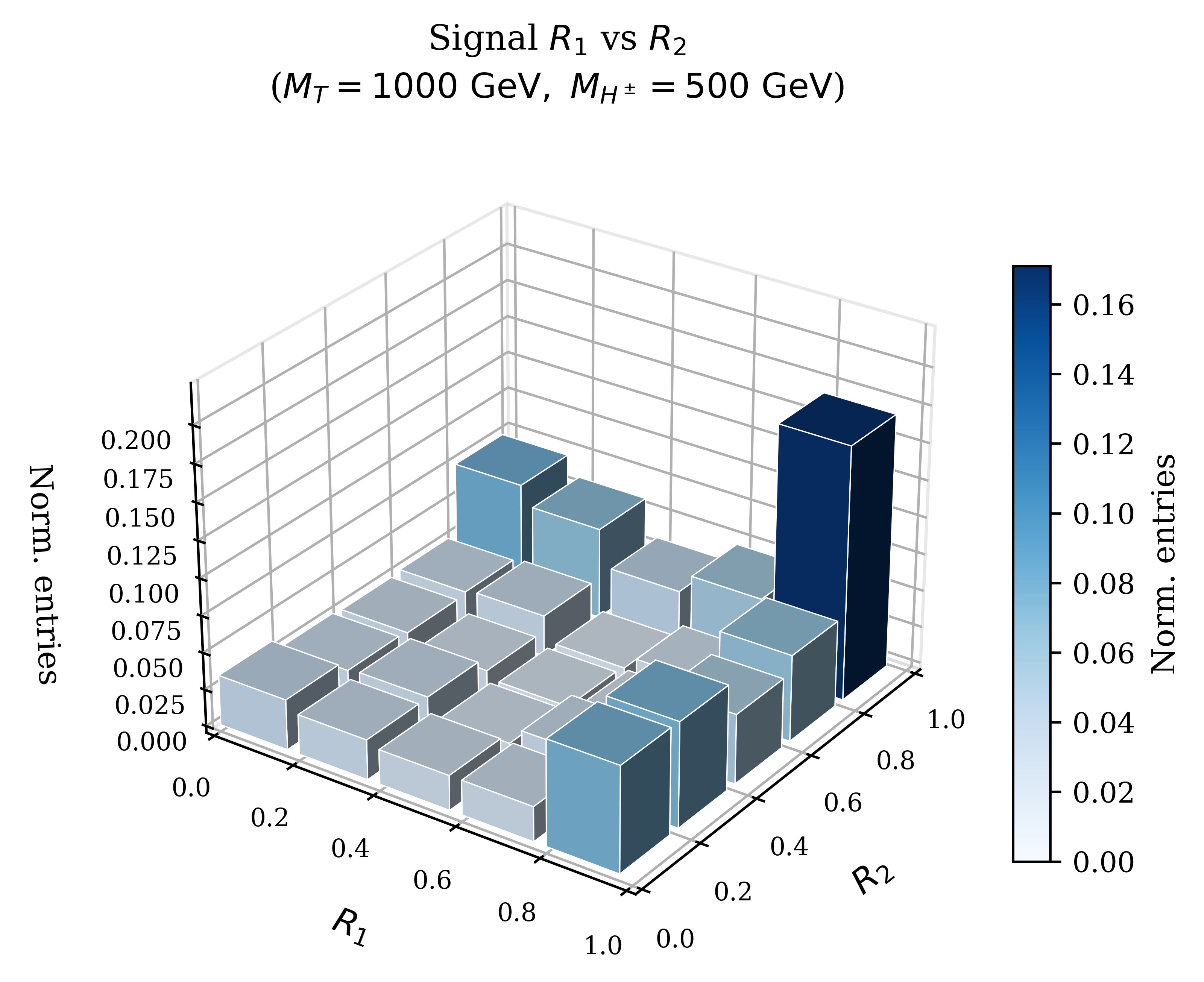}
\includegraphics[width=0.30\textwidth]{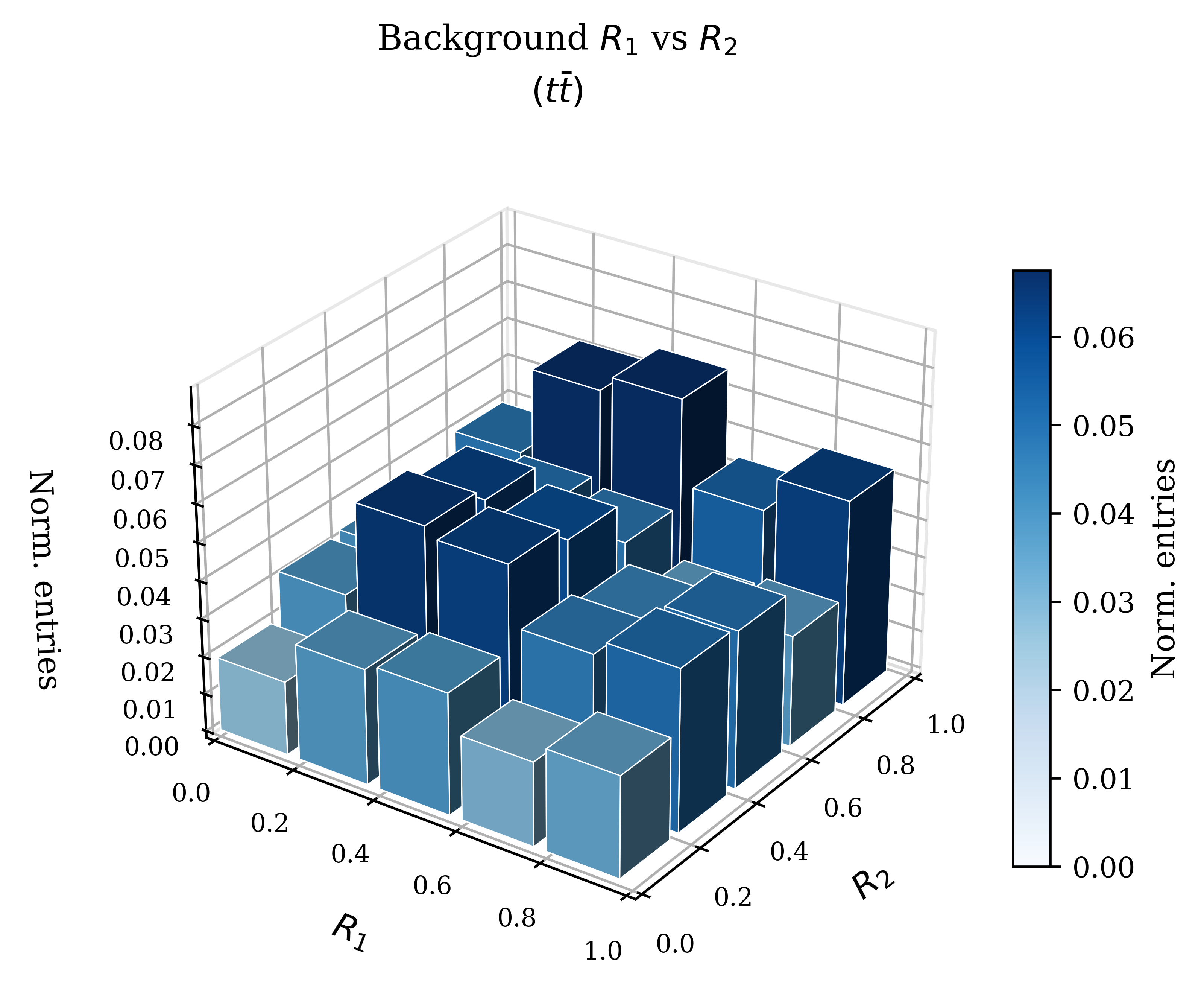}
\includegraphics[width=0.30\textwidth]{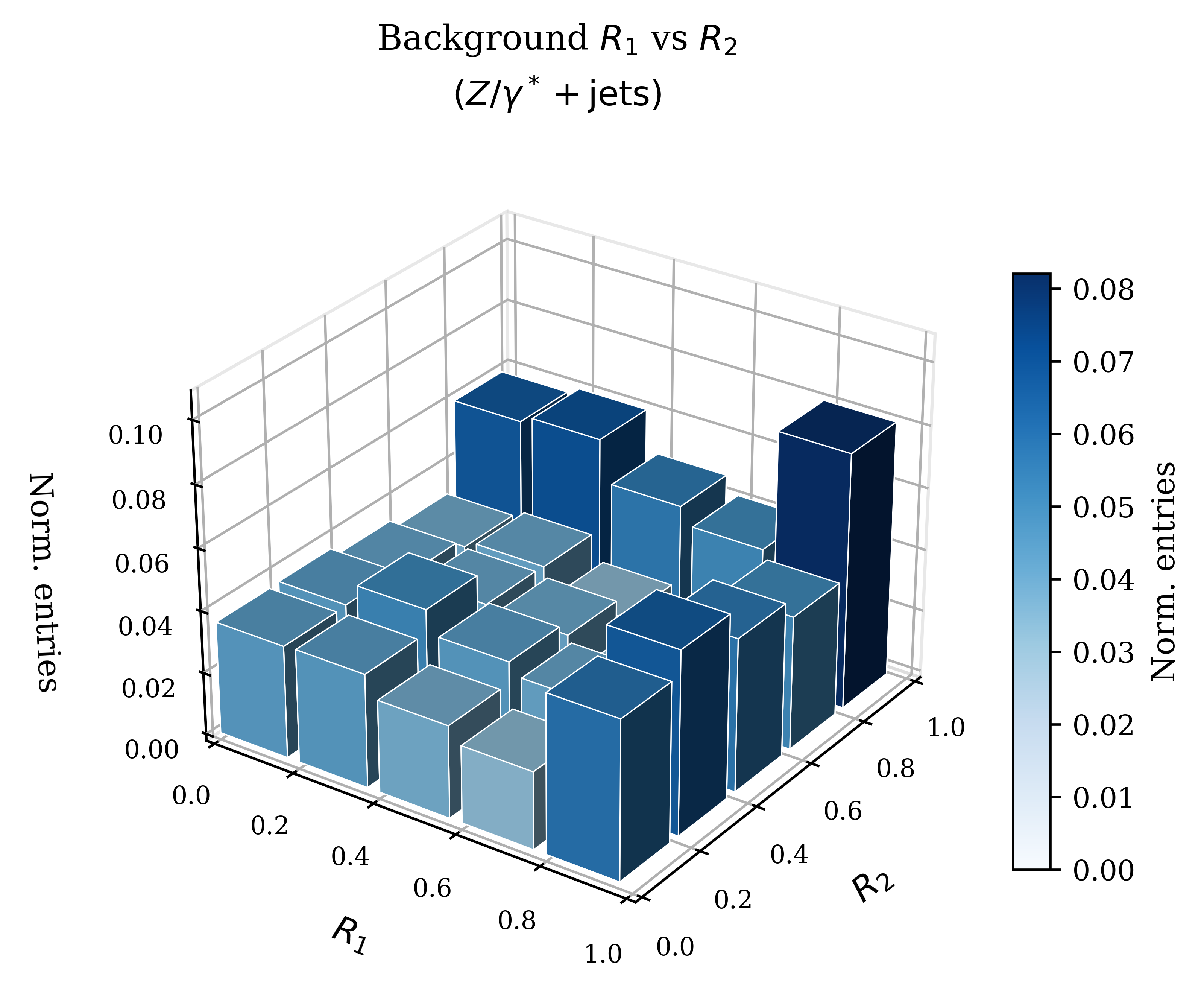}\\[6pt]
\includegraphics[width=0.30\textwidth]{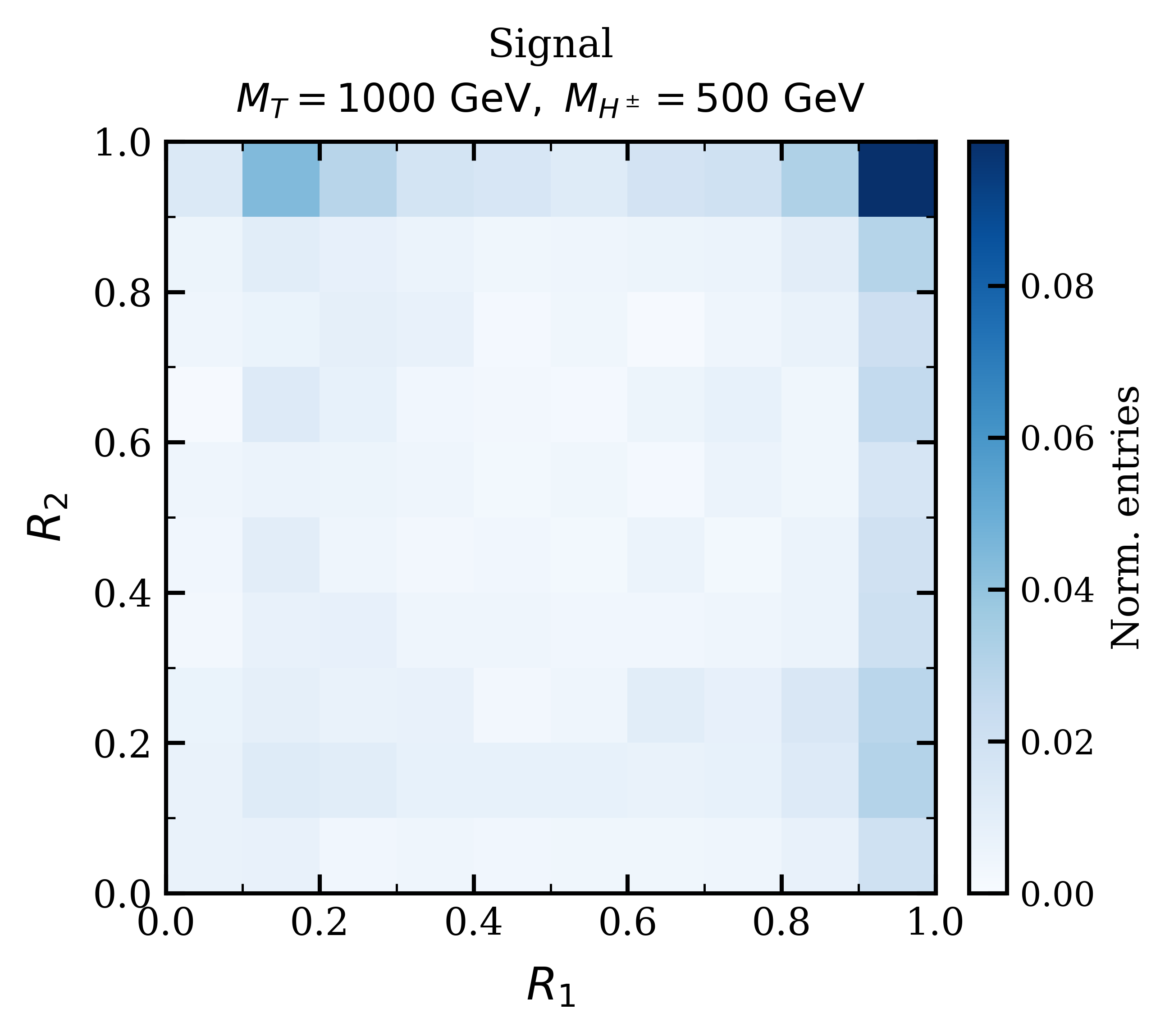}
\includegraphics[width=0.30\textwidth]{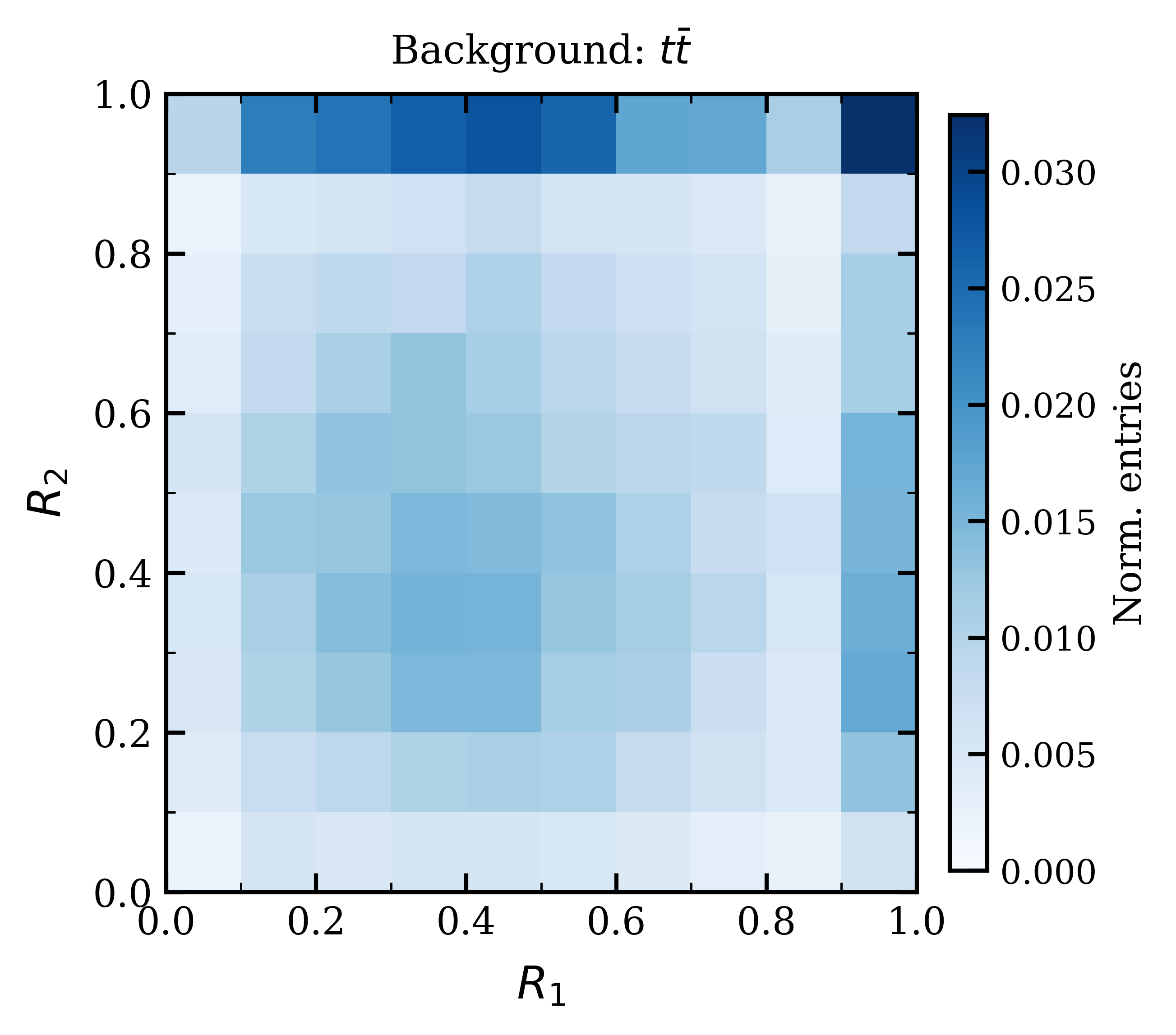}
\includegraphics[width=0.30\textwidth]{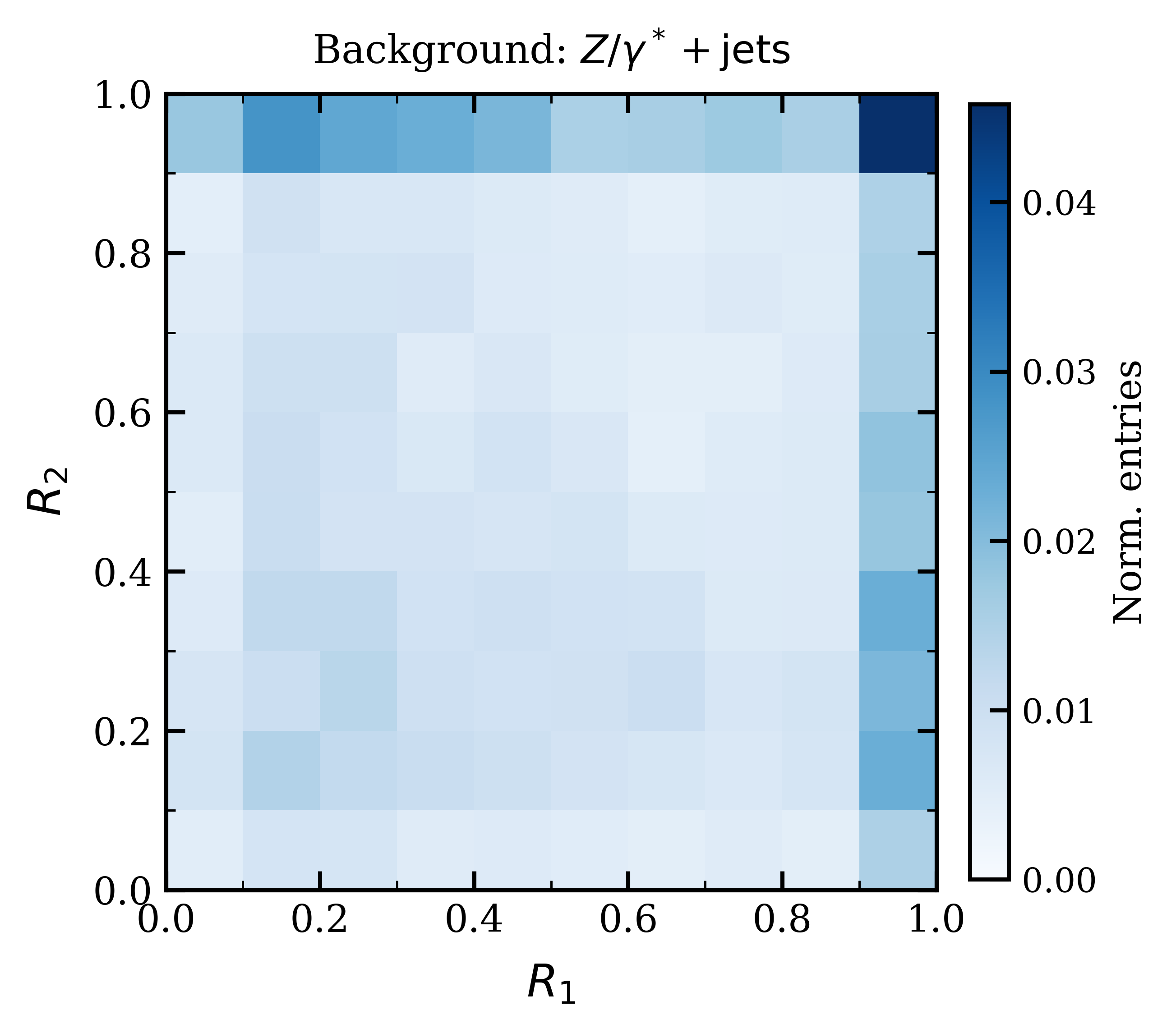}
\caption{Distributions of the tau polarization-sensitive variables
$R_1$ and $R_2$ for the signal ($M_T=1000$~GeV,
$M_{H^\pm}=500$~GeV), the $t\bar{t}$ background, and the
$Z(\to\tau\tau)+$jets background (left, centre, and right columns,
respectively). \textit{Top row}: three-dimensional histogram
representations. \textit{Bottom row}: corresponding 2D heatmaps
of normalized event fractions per bin. The signal accumulates at high
$R_1$, consistent with the right-handed tau polarization from scalar
$H^\pm$ decay, while the $t\bar{t}$ background populates the
low-$R_1$ region associated with left-handed taus from $W$ decays.
These observables are not included in the baseline cut-based
selection, but the clear separation demonstrated here motivates their
use in future multivariate or machine-learning analyses.}
\label{fig:R1R2}
\end{figure*}

The two-dimensional distributions in the $(R_1,R_2)$ plane for the
signal and the dominant backgrounds are shown in
Fig.~\ref{fig:R1R2}, in both 3D histograms and 2D
heatmap representations. The signal accumulates preferentially at
high values of $R_1$, consistent with the right-handed polarization
from scalar $H^\pm$ decay. The $t\bar{t}$ background, where taus are
produced via left-handed $W$ decays, is concentrated at low $R_1$.
The $Z(\to\tau\tau)+$jets background falls between the two, as the
$Z$ boson couples to both tau helicity states. The clear separation
visible in both representations confirms that $R_1$ and $R_2$ carry
discriminating power that is complementary to the kinematic
observables.

Although $R_1$ and $R_2$ are not incorporated into the cut-based
selection of Sec.~\ref{sec:analysis}, the clear separation visible
in Fig.~\ref{fig:R1R2} demonstrates that they carry independent
discriminating power. Their inclusion in a multivariate or
machine-learning analysis could yield a meaningful improvement in
discovery reach, particularly at large $M_T$ where the signal cross
section is small and kinematic discrimination alone may be
insufficient.

\section{Results}
\label{sec:results}

\subsection{Significance contours}

We present the expected discovery and exclusion reach of the HL-LHC
in the $2\tau+2b+\slashed{E}_T$ channel, evaluated from the Asimov
significance of Eq.~\eqref{eq:ZA} after applying the full event
selection of Table~\ref{tab:cutflow}. The event yields in
Table~\ref{tab:cutflow_events} correspond to two representative mass
points within this reach. Results are expressed as $2\sigma$
(exclusion) and $5\sigma$ (discovery) contours in two complementary
planes of the parameter space, shown in Fig.~\ref{fig:results}.

\subsection{The \texorpdfstring{$M_T$--$\mathcal{B}$}{MT-B} plane}

The left panel of Fig.~\ref{fig:results} shows significance contours
in the plane of $M_T$ versus the effective cascade decay rate 
$\mathcal{B}\equiv\mathrm{BR}(T\to H^\pm b)\times\mathrm{BR}(H^\pm\to\tau\nu)$,
for a fixed charged Higgs mass of $M_{H^\pm}=500$~GeV. The
light-blue band between the two contours identifies the parameter
region observable at $2\sigma$ but not yet at the $5\sigma$ discovery
threshold; the red region lying below the $2\sigma$ contour is not
accessible at this luminosity. Both contours terminate at $\mathcal{B}=1$,
which corresponds to the physical upper bound of the BR product.

The $5\sigma$ discovery contour reaches $M_T\approx1.9$~TeV for
$\mathcal{B}=1$, while the $2\sigma$ exclusion extends to
$M_T\approx2.05$~TeV. For moderate BRs, the required
$\mathcal{B}$ rises steeply with $M_T$ as the $T\bar{T}$ production
cross section falls, so the channel retains sensitivity primarily for
$M_T\lesssim1.9$~TeV at $5\sigma$ and $M_T\lesssim2.0$~TeV at
$2\sigma$.

\subsection{The \texorpdfstring{$M_{H^\pm}$--$M_T$}{MH-MT} mass plane}

\begin{figure*}[t!]
\centering
\includegraphics[width=0.46\textwidth]{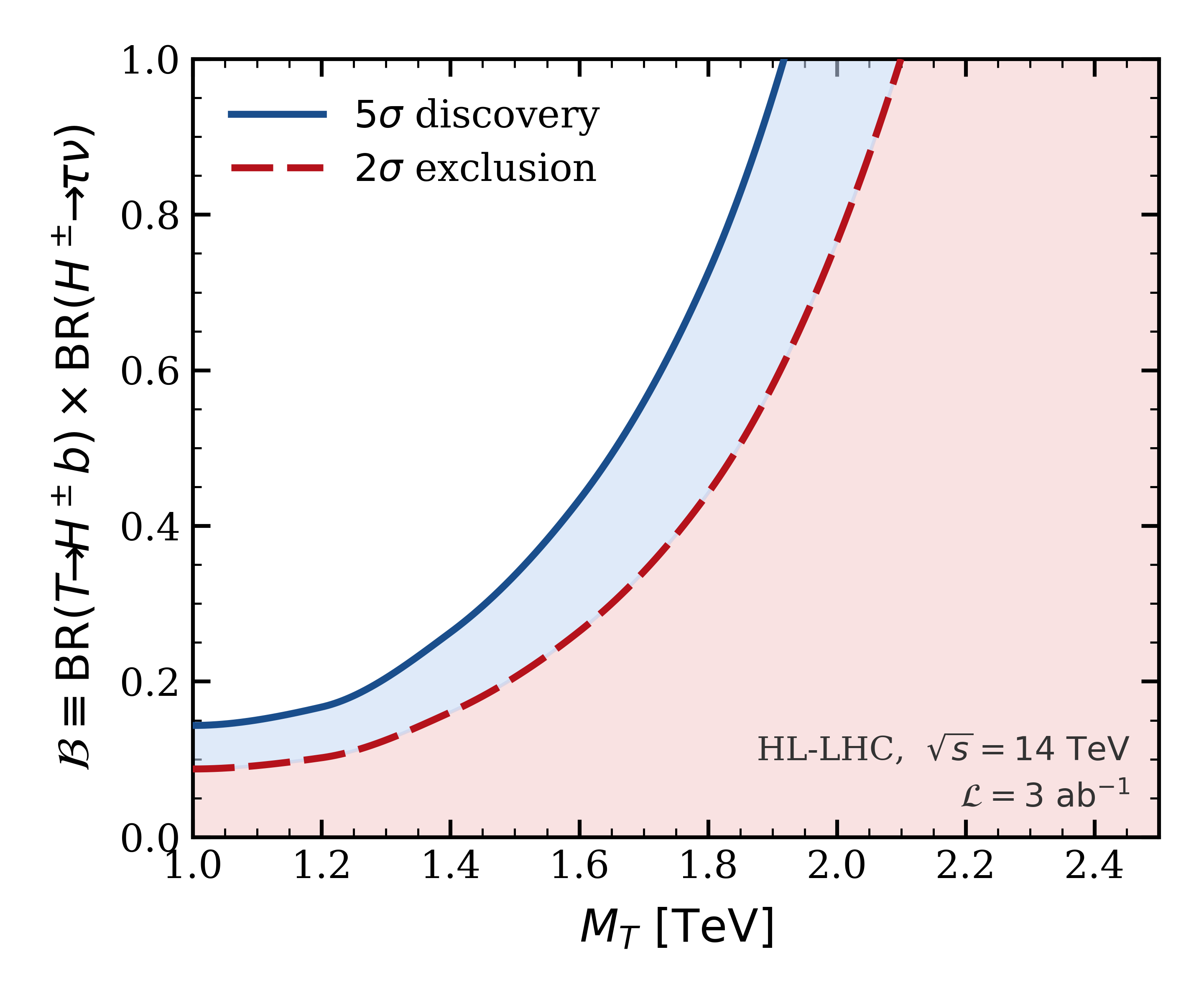}
\hfill
\includegraphics[width=0.46\textwidth]{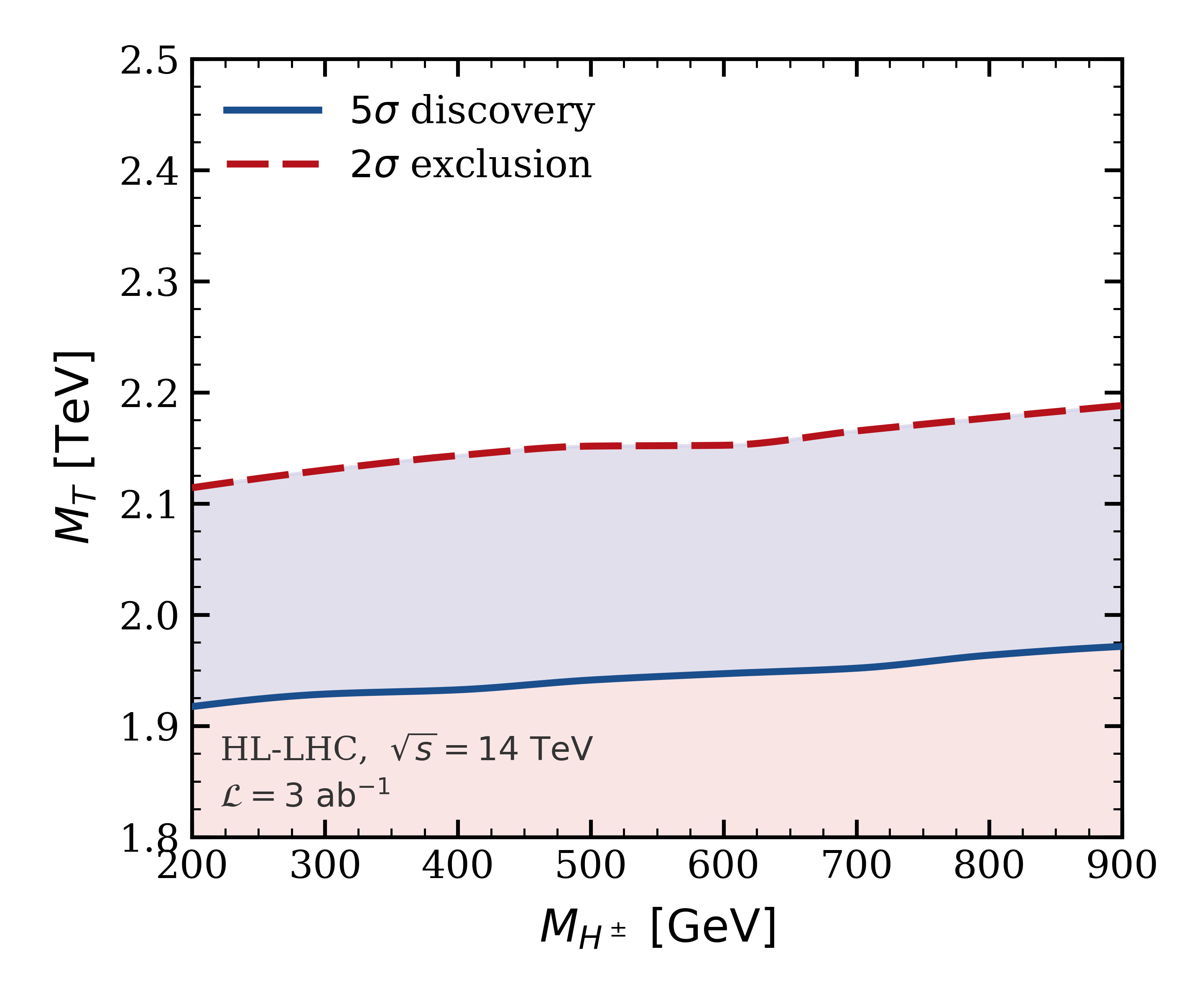}
\caption{Expected $2\sigma$ (dashed, red) and $5\sigma$ (solid, blue)
significance contours at the HL-LHC
($\mathcal{L}=3~\mathrm{ab}^{-1}$, $\sqrt{s}=14$~TeV) for the
$2\tau+2b+\slashed{E}_T$ channel.
\textit{Left}: Contours in the $M_T$--$\mathcal{B}$ plane with
$M_{H^\pm}=500$~GeV fixed, where
$\mathcal{B}\equiv\mathrm{BR}(T\to H^\pm b)\times\mathrm{BR}(H^\pm\to\tau\nu)$.
The blue shaded band denotes the region observable at $2\sigma$ but
below the $5\sigma$ discovery threshold; the red shaded region is
not significant at the $2\sigma$ level.
\textit{Right}: Contours in the $M_{H^\pm}$--$M_T$ mass plane for
$\mathcal{B}=1$, with the same shading convention.
A flat systematic uncertainty of $10\%$ on the total background
normalization is assumed throughout.}
\label{fig:results}
\end{figure*}

The right panel of Fig.~\ref{fig:results} shows the reach in the
$(M_{H^\pm},M_T)$ mass plane under the assumption $\mathcal{B}=1$,
representing the optimistic scenario in which the full signal rate
flows into the $2\tau+2b+\slashed{E}_T$ channel. The $5\sigma$ discovery
contour spans $M_T\approx1.90$--$1.97$~TeV across the full
$M_{H^\pm}$ range considered, while the $2\sigma$ exclusion reaches
$M_T\approx2.12$--$2.18$~TeV. The mild dependence on $M_{H^\pm}$
reflects the relatively flat signal efficiency as a function of the
charged Higgs mass once the kinematic selection requirements are
satisfied.

A modest reduction in sensitivity is observed near the kinematically
compressed region $M_T-M_{H^\pm}\sim m_b$, where the $b$-jet from
$T$ decay becomes insufficiently energetic to pass the $p_T$
threshold. This loss could be partially recovered through a dedicated
soft-$b$ selection strategy, which we leave as a direction for future
work.

\subsection{Discussion}

The results presented above demonstrate that the $2\tau+2b+\slashed{E}_T$
channel provides discovery-level sensitivity to non-standard VLQ
decays for $M_T\lesssim1.9$~TeV at the HL-LHC, assuming order-unity
cascade BRs. This reach is complementary to, and
largely orthogonal from, existing top--VLQ searches, which assume
exclusive decays into $Wb$, $Zt$, and $ht$, thereby not efficiently generating the
$2\tau+2b+\slashed{E}_T$ final state. The present analysis therefore
fills a significant gap in the existing top--VLQ search programme.

After the full event selection, the residual background is dominated
by $t\bar{t}$ production, which contributes the bulk of the surviving
noise, followed by $Z(\to\tau\tau)+$jets as a
sub-dominant but non-negligible contribution at lower $S_T$ values.
The inclusion of the tau polarization observables $R_1$ and $R_2$
discussed in Sec.~\ref{sec:pol} into a multivariate discriminant
would provide additional rejection of the $t\bar{t}$ background,
since signal and $t\bar{t}$ events populate opposite ends of the
$R_1$ spectrum. We leave also the development of such a multivariate
strategy as a direction for future work.

The assumed $10\%$ systematic uncertainty on the background
normalization is conservative relative to what could be achieved in a
full experimental analysis. In practice, dedicated control regions
enriched in $t\bar{t}$ (e.g.,~an $e\mu+\geq2b$ region) and in
$Z\to\tau\tau$ (e.g.,~a $Z\to ee$ region) can typically constrain
the background normalization to the few-percent level. Reducing
$\sigma_b/b$ from $10\%$ to $5\%$ would improve the $5\sigma$ mass
reach by approximately $50$--$100$~GeV.

\section{Conclusions}
\label{sec:conclusion}

We have investigated the prospects for observing non-standard decay
modes of VLQs partners to top-quarks at the HL-LHC, in the
context of a type-II 2HDM augmented by a singlet
VLQ. The signal topology
$pp\to T\bar{T}\to(H^+b)(H^-\bar{b})\to 2\tau+2b+\slashed{E}_T$
is distinctive, experimentally clean, and largely orthogonal to the
standard VLQ search channels targeting $Wb$, $Zt$, and $ht$ final
states.

A model-independent, cut-based analysis has been performed using
global kinematic observables constructed from the final-state visible
objects and the missing transverse momentum vector. The sequential
event selection—comprising requirements on $\slashed{E}_T$, the
scalar sum $S_T$, and several multi-body effective invariant masses
built from an auxiliary missing-momentum four-vector—suppresses the
dominant $t\bar{t}$ and $Z(\to\tau\tau)+$jets backgrounds by several
orders of magnitude while maintaining good signal efficiency across
the mass range considered. The results are expressed as
model-independent contours in the $M_T$--$\mathcal{B}$ and
$M_{H^\pm}$--$M_T$ planes, enabling direct reinterpretation in other
BSM scenarios with analogous cascade topologies.

Using the Asimov significance formalism with a conservative $10\%$
systematic uncertainty on the background normalization, we find that
the HL-LHC with $3~\mathrm{ab}^{-1}$ can achieve $5\sigma$ discovery
sensitivity for $M_T\lesssim1.9$~TeV, and can exclude at $2\sigma$
up to $M_T\approx2.1$~TeV, assuming
$\mathcal{B}=\mathrm{BR}(T\to H^\pm b)\times\mathrm{BR}(H^\pm\to\tau\nu)=1$.
These results establish the $2\tau+2b+\slashed{E}_T$ final state as a
well-motivated and sensitive probe of non-standard top--VLQ decays at the
HL-LHC.

We have also shown that the tau polarization variables $R_1$ and
$R_2$, which encode the helicity of the parent particle through the
kinematics of hadronic tau decay products, provide a natural
complement to the kinematic event selection. The opposite
polarizations of signal taus (from $H^\pm$) and dominant background
taus (from $W^\pm$) manifest as a clear separation in the $(R_1,R_2)$
plane (Fig.~\ref{fig:R1R2}), offering additional discriminating power
that could be harnessed in a future multivariate or machine-learning
analysis to extend the discovery reach beyond what is achievable with
the cut-based approach alone.

Taken together, these findings motivate dedicated experimental
searches for non-standard top--VLQ decay modes at the HL-LHC and
underscore the importance of considering extended Higgs sector
signatures when interpreting the results of VLQ searches, in the spirit of
Refs.~\cite{Benbrik:2022kpo,Arhrib:2024tzm}.

\acknowledgments
T.M. acknowledges partial support from the SERB/ANRF,
Government of India, through the Core Research Grant
(CRG) No. CRG/2023/007031. 
S.M. is supported in part through the NExT Institute and STFC Consolidated Grant ST/X000583/1. 
R.S. acknowledges the Prime Minister's Research Fellowship (PMRF ID: 0802000).

\bibliography{References}
\bibliographystyle{JHEPCust}

\end{document}